\documentclass{ecai}

\usepackage{latexsym}
\usepackage[utf8]{inputenc}
\usepackage{amssymb}
\usepackage{bm}
\usepackage{dsfont}
\usepackage{nccmath}
\usepackage{amsmath}
\usepackage{amsthm}
\usepackage{amsfonts}
\usepackage{algorithm}
\usepackage{algorithmic}
\usepackage{graphicx}
\usepackage{multirow}
\usepackage{diagbox}
\usepackage[export]{adjustbox}
\usepackage{comment}
\usepackage{color}
\usepackage{subfigure}
\usepackage{caption}
\usepackage{enumitem}
\usepackage{mathtools}
\usepackage{physics}

\newtheorem{lemma}{Lemma}
\newtheorem{theorem}{Theorem}

\newtheorem{problem}{Problem}
\newtheorem{remark}{Remark}
\newtheorem{notations}{Notations}

\DeclareMathOperator*{\argmax}{arg\,max}

\begin{document}
\begin{frontmatter}
\paperid{993}

\title{Dirichlet Logistic Gaussian Processes\\ for Evaluation of Black-Box Stochastic Systems\\ under Complex Requirements}

\author[A]{\fnms{Ryohei}~\snm{Oura}\orcid{0000-0001-9864-4506}\thanks{Corresponding Author. Email: ryohei\_oura@mail.toyota.co.jp}}
\author[B]{\fnms{Yuji}~\snm{Ito}\orcid{0000-0002-6786-7798}}

\address[A]{Frontier Research Center, Toyota Motor Corporation, Japan}
\address[B]{Toyota Central R\&D Labs., Inc., Japan}


\begin{abstract}
    The requirement-driven performance evaluation of a black-box cyber-physical system (CPS) that utilizes machine learning methods has proven to be an effective way to assess the quality of the CPS. However, the distributional evaluation of the performance has been poorly considered.
    Although many uncertainty estimation methods have been advocated, they have not successfully estimated highly complex performance distributions under small data.
    In this paper, we propose a method to distributionally evaluate the performance under complex requirements using small input-trajectory data. To handle the unknown complex probability distributions under small data, we discretize the corresponding performance measure, yielding a discrete random process over an input region. Then, we propose a semiparametric Bayesian model of the discrete process based on a Dirichlet random field whose parameter function is represented by multiple logistic Gaussian processes (LGPs).
    The Dirichlet posterior parameter function is estimated through the LGP posteriors in a reasonable and conservative fashion.
    We show that the proposed Bayesian model converges to the true discrete random process as the number of data becomes large enough.
    We also empirically demonstrate the effectiveness of the proposed method by simulation.
\end{abstract}

\end{frontmatter}

\section{Introduction}
Evaluation of the quality of cyber-physical systems (CPSs) with complex behaviors for given requirements is vital since the major applications of the CPSs are safety-critical. The quality evaluation is challenging because CPSs involve stochastic uncertainty and industry-scale CPSs make the requirements complex. Complex requirements for systems are usually translated into real-valued performance measures (robustness degrees). For example, a signal temporal logic (STL) formula formally defines how robustly trajectories of system variables generated under inputs satisfy requirements \cite{donze2010robust, fainekos2009robustness}. Such performance measures are difficult to evaluate owing to the stochasticity of CPSs.

Although machine learning-based methods \cite{deshmukh2017testing, mathesen2019falsification, mathesen2021efficient, pedrielli2021part, humeniuk2022search, ben2016testing, haq2022efficient, fan2020statistical}
have been actively proposed for evaluating requirements, they often overlook risks that requirements are not satisfied. The methods have represented robustness degrees by surrogate models such as Gaussian processes (GPs) with sample trajectories of system variables. Confidence of the surrogate models has been quantified to judge whether the sample trajectories are sufficient for the evaluation \cite{deshmukh2017testing, mathesen2019falsification, mathesen2021efficient, pedrielli2021part, fan2020statistical}.
Unfortunately, the robustness degrees are simply evaluated as deterministic values or their expectations. Such deterministic or expected evaluation neglects cases where the robustness degrees have exceptionally poor values with low probability.

To assess the aforementioned risks owing to the stochasticity of systems, it is important to evaluate the distributions of robustness values. Some works \cite{le2005heteroscedastic, makarova2021risk} have harnessed input-dependent sub-Gaussian noise with GPs. They, however, cannot derive the probability distribution itself. In  \cite{skovbekk2023formal, yu2023discretization}, the authors have employed the discretization of the target value to handle the complexity of the corresponding probability distribution. They have shown that such discretization achieves a more robust and accurate estimation than directly treating the target value.
In addition, such distributional evaluation should be realized using a small data set because obtaining data samples from high-fidelity simulators or real-world experiments is expensive and time-consuming.

Quantifying the estimation uncertainty is important to identify whether the estimated distribution is reliable.
While a variety of uncertainty quantification methods for discrete distributions have been advocated \cite{milios2018dirichlet, liu2021scalable, sensoy2018evidential, charpentier2020posterior, joo2020being, shen2023post}, they have faced a critical challenge: the accurate estimation and appropriate uncertainty quantification under a small data set.
In \cite{milios2018dirichlet, liu2021scalable}, the authors have considered estimating a discrete distribution for any input by latent GPs. Specifically, the authors in \cite{milios2018dirichlet} have used a Dirichlet prior and approximately regressed its parameter vector through moment matching and multiple GPs.
However, the uncertainty of the estimated distribution obtained from \cite{milios2018dirichlet, liu2021scalable} is determined depending on the kernel functions rather than the number of data points, which yields the possibility that the estimation is overconfident (or underconfident).
In \cite{sensoy2018evidential, charpentier2020posterior, joo2020being, shen2023post}, the authors have quantified the estimation uncertainties by placing a Dirichlet distribution over a probability simplex. Based on density estimation and direct posterior matching, they have obtained a reasonable posterior Dirichlet distribution by calculating the pseudo-count of observations of each class at any input. Although the posterior calculation based on the pseudo observation counts is reasonable in the sense of Bayesian updates, the methods have the following two issues. (a) They require huge data to adopt gradient-based learning of deep neural networks. (b) Moreover, they cannot handle the estimation uncertainty of the pseudo-counts themselves. In the context of the evaluation of safety-critical CPS, it is desirable to suppress an overconfident and biased estimation under small data.

Logistic Gaussian processes (LGPs) can be employed to overcome the aforementioned issues (a) and (b). LGPs are well-known Bayesian nonparametric priors for unknown density functions \cite{leonard1978density, tokdar2007towards, riihimaki2014laplace, donner2018efficient,  tokdar2007posterior, adams2009tractable, donner2018efficient_gcp, tokdar2022heavy}. 
In \cite{tokdar2007towards, riihimaki2014laplace, donner2018efficient}, the authors have proposed tractable inference methods for the LGP posteriors. In \cite{tokdar2007posterior}, the authors have shown the theoretical aspects of LGPs such as posterior consistency.
Utilizing LGP priors, we will quantify the confidences of density-based pseudo-counts according to the amount of data without specific model parameters. 

In this paper, for a black-box stochastic system equipped with the robustness degree that describes a complex requirement, we propose a semiparametric Bayesian estimation method for the probability distribution of the robustness values over a compact input region. The proposed method is mainly based on two ingredients to overcome the aforementioned challenge. First, we consider the discretization of the robustness degree to treat the complex probability distribution under a small data set. Second, we consider placing a Dirichlet random field on probability simplices of discretized robustness values and utilizing multiple LGPs to estimate the Dirichlet posterior. Moreover, this approach simultaneously considers the suppression of overconfidence and the goodness-of-fit to the data, taking into account the amount of data.

Our contributions are as follows:
\begin{enumerate}
    \item We propose a semiparametric model of a discrete random process based on a Dirichlet random field and multiple LGPs. This model quantifies the uncertainty of the Dirichlet posterior parameter function itself.
    \item By introducing a conservativeness parameter, we directly adjust the suppression of the overconfidence and the goodness-of-fit to the data according to the amount of data. The parameter determines the extent to which the uncertainty of LGP posteriors is incorporated into the pseudo-counts of observations.
    \item We show that the posterior of the proposed model converges to the true one as the data number goes to infinity under mild conditions.
    \item We demonstrate the effectiveness of the proposed method by a path planning problem and compare it with existing estimation methods. The proposed method is more accurate than the existing methods and captures the areas with few data points through the estimation uncertainty.
\end{enumerate}

The rest of the paper is organized as follows. 
Section \ref{problem_setup} formulates a problem of estimating the discrete random process that respects a black-box stochastic system with a robustness degree from small input-trajectory data. 
Section \ref{Modeling_Estimation} proposes a semiparametric Bayesian estimation method based on a Dirichlet random field and multiple LGPs. 
Section \ref{Analysis} analyzes the convergence property of the proposed model. 
Section \ref{Example} gives a numerical example to demonstrate the effectiveness of our proposed method.

\begin{notations}
$\mathbb{R}$, $\mathbb{R}_{\geq 0}$, and $\mathbb{R}_{> 0}$ are the set of real, non-negative real, positive real numbers, respectively. We denote the cardinality of a set $T$ by $|T|$. For any $m \in \mathbb{N}$, we denote $[m] = \{ 1, \ldots, m \}$.
For any $K \in \mathbb{N}$, $i$-th entry of a vector $\bm{v} \in \mathbb{R}^K$ and $i,j$-th entry of a matrix $G \in \mathbb{R}^{K\times K}$ are represented by the subscripted character $v_i$ and $G_{ij}$, respectively.
\end{notations}


\section{Problem setup}
\label{problem_setup}
We consider a (black-box) stochastic system that generates a trajectory $y: \mathbb{R}_{\geq 0} \to \mathbb{R}^o$ randomly from an input $x$ in $\mathcal{X}$, where $\mathcal{X} \subset \mathbb{R}^d$ is a closed set. For each $x$, the probability distribution over $\mathcal{Y}$ is denoted by unknown $\mathcal{S}(y|x)$, where $\mathcal{Y}$ is the set of trajectories $y$.
Let $\rho: \mathcal{Y} \to \mathbb{R}$ be a predefined robustness degree (performance measure) of each trajectory $y$.
For example, using STL formulas defines the robustness degrees in a successful and formal manner from a given complex requirement \cite{fainekos2009robustness, donze2010robust}.

The robustness degree is distributionally estimated using a small data set.
Let $X:= \{ x_i \}_{i=1}^N \subset \mathcal{X}$ and $Y:= \{ y_i \}_{i=1}^N \subset \mathcal{Y}$ be the input and trajectory data, respectively, where the input data are independent and identically distributed (i.i.d.) from a given probability distribution $p(x)$ and each $y_i$ is generated from $\mathcal{S}(\cdot | x_i)$. 

Before estimating the probability distribution of $\rho(y)$, inspired by some recent works \cite{skovbekk2023formal, yu2023discretization}, we classify $\rho(y)$ into $m$ robustness levels for a given $m \in N$. Such classification makes the estimation feasible even when the data set size is small and the distribution of $\rho(y)$ is highly complex.
The levels are defined by disjoint intervals $L_{l}$ such that $\cup_{l=1}^{m} L_{l}=\mathbb{R}$ holds.
For each $x \in \mathcal{X}$, the probabilities ${\pi}_{l}(x)$ of the robustness degree $\rho(y)$ falling within $m$ robustness levels are given by
\begin{align}
\label{discretized_process}
{\pi}_{l}(x) 
:=\mathrm{Pr}(  \rho(y) \in L_{l}  | x ).
\end{align}

The goal of this study is to estimate the probability vector $\bm{\pi}(x)$ using the given input-trajectory data $(X, Y)$, where we engage in the case that $|X|$ is small.
Then, we consider the following problem.

\begin{problem}
\label{problem}
	Given an input-trajectory data $(X, Y)$ obeying a black-box stochastic system, a robustness degree $\rho$, and $m$ robustness levels $L_1,\ldots, L_m$,
	estimate a provability vector $\bm{\pi}: \mathcal{X} \to [0,1]^{m}$ of the robustness degree $\rho(y)$ falling within $m$ robustness levels in the senses of the mean and confidence bounds.
\end{problem}

\begin{remark}
    We face Problem \ref{problem} in the performance evaluation of CPSs in a data-driven manner. For example, let us consider the path planning performance of a robot in an environment with obstacles. Let x and y denote the starting position and trajectory of the robot, respectively. The trajectory contains randomness owing to the stochastic dynamics of the robot. A robustness degree $\rho(y)$ evaluates the path planning performance for each trajectory $y$. It is crucial to evaluate the probability vector $\pi$ using small data of (X,Y) so that the performance of black-box systems is efficiently evaluated. Such an example is demonstrated in Section \ref{Example}.
\end{remark}

\section{Dirichlet Logistic Gaussian Process}
\label{Modeling_Estimation}
In this section, we propose a semiparametric Bayesian model called the Dirichlet logistic Gaussian process (\textsf{DLGP}) to solve Problem 1 that involves a discrete random process. 
In the \textsf{DLGP}, we reduce solving Problem \ref{problem} into estimating a posterior Dirichlet random field. The posterior parameter function of the Dirichlet random field is represented by multiple LGPs. The function is estimated to balance the reduction of the overconfidence and the goodness-of-fit to the data.
We explain the modeling and computation of the \textsf{DLGP} in subsections \ref{section:Modeling} and \ref{section:alpha_post}, respectively.

\subsection{Bayesian Semiparametric Discrete Process Modeling}
\label{section:Modeling}
The proposed \textsf{DLGP} provides an estimate $\hat{\bm{\pi}}$ of the probability vector $\bm{\pi}: \mathcal{X} \to [0,1]^{m}$ based on a Dirichlet random field $Dir( \hat{\bm{\pi}}(\cdot) | \bm{\alpha}_\textsf{post}(\cdot))$ as follows \cite{charpentier2020posterior, miller2023dirichlet}:
\begin{align}
\label{Dirichlet_random_field}
    \hat{\bm{\pi}}(x) \sim Dir(\cdot | \bm{\alpha}_\textsf{post}(x)).
\end{align}
Note that the Dirichlet random field offers a Dirichlet distribution \cite{gelman2013bayesian} defined by $Dir(\hat{\bm{\pi}}(x) | \bm{\alpha}_\textsf{post}(x)) \propto \prod_{i=1}^m \hat{\pi}_i(x)^{\alpha_{\textsf{post},i}(x) - 1}$, where $\bm{\alpha}_\textsf{post}(x) \in \mathbb{R}_{\geq 0}^m$ is a posterior parameter function determined later. 
The representation of $\hat{\bm{\pi}}$ provides the capability of acquiring the appropriate estimator and the estimation confidence.

If the posterior parameter function ${\bm{\alpha}}_\textsf{post}(x)$ of the Dirichlet random field is estimated as $\hat{\bm{\alpha}}_\textsf{post}(x)$, $Dir(\cdot|\hat{\bm{\alpha}}_\textsf{post})$ provides our solution to Problem 1. Namely, the mean and confidence bounds of $\bm{\pi}$ are calculated from $Dir(\cdot | \hat{\bm{\alpha}}_\mathsf{post})$.


We place a Dirichlet prior $Dir(\cdot|\bm{\alpha}_{\textsf{prior}})$ over the probability vector $\bm{\pi}(x)$ for any $x$, where $\bm{\alpha}_{\textsf{prior}} \in \mathbb{R}^m$ is a prior parameter vector. 
Let $N_l := | \{ y \in Y \;|\; \rho(y) \in L_{l}  \} |$. When $\mathcal{X}$ is a singleton, that is $\mathcal{X} = \{ x \}$, 
the $i$-th posterior parameter is defined as $\alpha_{\textsf{post}, l}(x) = N_l + \alpha_{\textsf{prior}, l}$ \cite{gelman2013bayesian}. Thus,  a reasonable value of the posterior parameter vector at $x$ is provided as
\begin{align}
\label{post_param_func}
    \bm{\alpha}_{\textsf{post}}(x) = \bm{\alpha}(x) + \bm{\alpha}_{\textsf{prior}},
\end{align}
where the \textit{pseudo-count} function $\bm{\alpha}: \mathcal{X} \to \mathbb{R}_{\geq 0}$ is defined as 
\begin{align}
\label{ideal_pseudo_count}
    \alpha_l(x) = N_l p(x | l)
\end{align}
and $p(x | l)$ denotes a probability density of $x$ in $\mathcal{X}$ under the robustness degree classified in the $l$-th robustness level $L_l$. Intuitively, $\alpha_l(x)$ in (\ref{ideal_pseudo_count}) indicates the frequency of data points whose robustness degrees lie in $L_l$ at $x$.

We represent the density function $p(x|l)$ through a latent GP prior. A GP is a kernel method to define a distribution of a function $f$, which is dented by $f \sim \mathcal{GP}(m, g)$, where $m: \mathcal{X} \to \mathbb{R}$ and $g: \mathcal{X} \times \mathcal{X} \to \mathbb{R}$ are a mean function and a covariance kernel, respectively. This means that, for any set of inputs $\{ x_i \}_{i=1}^N$, the collection of the values $\bm{f} = ( f(x_1), \ldots, f(x_N) )$ has a joint Gaussian distribution as $\bm{f} \sim \mathcal{N}(\bm{m}, G)$, where the mean $\bm{m} \in \mathbb{R}^N$ and the covariance matrix $G \in \mathbb{R}^{N\times N}$ are defined as $m_i = m(x_i)$ and $G_{ij} = g(x_i,x_j)$, respectively.
We denote $g(\cdot, \cdot|\bm{\theta})$ when we explicitly indicate the kernel $g$ has the hyper-parameters $\bm{\theta} = \{ \theta_1, \ldots, \theta_n\}$.
We assume that the density function $p(\cdot|l)$ on $\mathcal{X}$ is represented as follows.
\begin{align}
\label{logistic_trandoform}
    p(x|l) = \frac{\exp(f_l(x))}{\int_{\mathcal{X}} \exp(f_l(s)) \mathrm{d}s},
\end{align}
where $f_l$ follows a zero-mean GP prior, namely, $f_l \sim \mathcal{GP}(\bm{0}, g_l)$.
Clearly, $p(\cdot|l)$ in (\ref{logistic_trandoform}) defines a stochastic process on $\mathcal{X}$ whose sample paths satisfy $p(x|l) \geq 0$ for any $x$ in $\mathcal{X}$ and $\int_{\mathcal{X}} p(x|l) \mathrm{d}x = 1$.
Then, we say that $p(\cdot|l)$ follows an LGP prior \cite{tokdar2007posterior} with a hyper-prior $H$, represented as follows.
\begin{align}
    p(\cdot | l) &\sim \mathcal{LGP}(\bm{0}, g_{l}(\cdot, \cdot |\bm{\theta})), \\
    \label{hyperparam_prior}
    \bm{\theta} &\sim H.
\end{align}
Thanks to the Bayesian nonparametric modeling, the estimator of $p(\cdot | l)$ can approximate the input-sample distribution well and has moderate smoothness simultaneously. Moreover, we can control the confidence of the density estimation by the LGP posterior. This leads to avoiding over-confidence in the pseudo-count.

Then, we define the estimator of the pseudo-count function $\bm{\alpha}$ as follows:
\begin{align}
\label{cons_pseudo_count}
    \hat{\alpha}_l(x;\lambda) = \max \{ N_l \left( p_\mathrm{E}(x | l, X) - \lambda p_\sigma(x | l. X) \right), 0 \}
\end{align}
by replacing $p(x|l)$ in (\ref{ideal_pseudo_count}) with $p_E(x|l, X) - \lambda p_\sigma(x|l, X)$,
where $p_\mathrm{E}(\cdot|l, X)$ and $p_\sigma(\cdot|l, X)$ are the mean and the standard deviation of the $l$-th LGP posterior, respectively, and $\lambda \in \mathbb{R}_{\geq 0}$ is a free parameter. Intuitively, $\hat{\alpha}_l(x;\lambda)$ represents a ``weakly believed value" of the pseudo-count at $x$. The parameter $\lambda$ determines how much we incorporate the lack of confidence of the density at $x$ into the pseudo-counts. So, we call $\lambda$ a conservativeness parameter. This directly enables us to balance the reduction of overconfidence and the goodness-of-fit to the data.

We obtain the estimator $\hat{\bm{\alpha}}_\textsf{post}$ of the posterior parameter function by substituting $\hat{\alpha}_l$ into (\ref{post_param_func}). That is, 
\begin{align}
\label{alpha_hat_post}
    \hat{\bm{\alpha}}_{\textsf{post}}(x; \lambda) = \hat{\bm{\alpha}}(x; \lambda) + \bm{\alpha}_{\textsf{prior}}.
\end{align}

\begin{remark}[Summary]
We summarize the intuitive interpretation of the overall discrete process model. First, the Dirichlet random field captures the belief of the discrete process about the robustness levels. Its prior parameter vector reflects the number of ``prior observations" for each robustness level under any input. Then, the posterior parameter function is defined through the pseudo-count function as an analogy to the Bayesian update of a Dirichlet distribution. The belief for the pseudo-count function is represented by the LGP prior.
\end{remark}


\subsection{Computation of Dirichlet Logistic Gaussian Process}
\label{section:alpha_post}
In the following, we mainly focus on computing the $m$ LGP posteriors and determining the conservativeness parameter.
First of all, we show the overall procedure of estimating the posterior parameter function of the Dirichlet random field through computing LGP posteriors and determining the conservativeness parameter in Algorithm \ref{Alg_PGDP}. From Line 1 to 3, Algorithm \ref{Alg_PGDP} proceeds by computing the LGP posterior for each $l$. In principle, we can apply any existing computation method for LGP posteriors \cite{riihimaki2014laplace, tokdar2007towards, donner2018efficient}. We use a discrete approximation method with a Markov chain Monte Carlo (MCMC) method in Section \ref{Example}.
In Line 4, we determine the conservative parameter $\lambda$ by (\ref{opt_lambda}) to balance the goodness-of-fit to the data and the conservativeness based on the amount of data. 
By calculating the estimator of the posterior parameter function in Line 5, we obtain the posterior Dirichlet random field $Dir(\cdot | \hat{\bm{\alpha}}_{\textsf{post}}(x; \lambda))$.

\begin{algorithm}[htb]
\caption{Dirichlet Logistic Gaussian Process}
\begin{algorithmic}[1] 
\label{Alg_PGDP}
\REQUIRE $(X, Y)$, $\rho$, $L_1, \ldots, L_m$, $q(\lambda)$, $H$.
\ENSURE $Dir(\cdot | \hat{\bm{\alpha}}_\textsf{post}(\cdot; \lambda^*))$.
\FOR{$l=1\ldots,m$}
\STATE Compute $l$-th LGP posterior using $X_l = \{ x_i \in X \;|\; y_i \in Y, \rho(y_i) \in L_l \}$  through (\ref{LGP_posterior}) and (\ref{hyper_param_posterior}).
\ENDFOR
\STATE Optimize the conservative parameter $\lambda$ by (\ref{opt_lambda}).
\STATE Obtain $\hat{\bm{\alpha}}_\textsf{post}(x; \lambda^*)$ in (\ref{alpha_hat_post}) from the computed LGP posteriors and the optimized $\lambda^*$.
\end{algorithmic}
\end{algorithm}

\subsubsection{Computaion of Logistic Gaussian Process Posteriors}
In what follows, we review a computation method for the posteriors of each LGP and its hiper-parameters based on a discrete approximation by dividing $\mathcal{X}$ into grids \cite{riihimaki2014laplace} employed in Section \ref{Example}.  
First, we divide the input region $\mathcal{X}$ into the set of sub-regions $S = \{ s_i \}_i$.
For simplicity, we assume that each sub-region is the same size. We denote $Z$ as the set of the centers of sub-regions.
Then, we consider computing each $l$-th LGP posterior on $S$. Let $X_l$ be the subset of $X$ whose corresponding trajectories have robustness values falling within $L_l$.
According to (\ref{logistic_trandoform}), the likelihood contribution of an input in $X_l$ that belongs to $i$-th grid is written as
\begin{align}
    \ell_i = \frac{\exp(f_{l,i})}{\sum_k \exp(f_{l,k})},
\end{align}
where the latent value $f_{l,i}$ corresponds to the $i$-th grid. We represent the set of $f_{l,i}$ for all grids as $F_l$. We denote the number of input samples in $X_l$ that fall within the $i$-th grid as $c_{l,i}$ and the set of them as $C_l$. The entire likelihood contribution of $X_l$ is given by
\begin{align}
\label{logistic_likelihood_contr}
    p(C_l | F_l) = \frac{\prod_{i=1}^{|S|} \exp(f_{l,k})^{c_{l,i}} }{ (\sum_k \exp(f_{l,k}))^{|X_l|} }.
\end{align}
Using the zero-mean GP prior $p( F_l | Z, \bm{\theta}_l)$, the prior on the hyper-parameter $p(\bm{\theta}_l)$ in (\ref{hyperparam_prior}), and the logistic likelihood contribution in (\ref{logistic_likelihood_contr}), we obtain the following posterior by Bayes' rule:
\begin{align}
\label{LGP_posterior}
    p(F_l | Z, C_l, \bm{\theta}_l) \propto p(C_l | F_l) p(F_l| Z, \bm{\theta}_l).
\end{align}
Likewise, we obtain the posterior of the hyper-parameters as
\begin{align}
\label{hyper_param_posterior}
    p(\bm{\theta}_l | Z, C_l, F_l) \propto p(C_l | F_l) p(F_l| Z, \bm{\theta}_l) p(\bm{\theta}_l).
\end{align}
We can compute each posterior by using any Markov chain Monte Carlo method \cite{gelman2013bayesian} such as the no-U-turn sampler \cite{hoffman2014no} and any approximation inference such as Laplace approximation \cite{riihimaki2014laplace}.

\subsubsection{Optimization of Conservativeness Parameter}
We choose an optimal conservativeness parameter $\lambda$ in (\ref{cons_pseudo_count}) by introducing a prior, $\lambda \sim q(\cdot)$. The prior denotes a priori requirement on how conservative the pseudo-count estimates should be. Then, we determine an optimal parameter $\lambda^*$ by maximum a posteriori (MAP) estimation as follows.
\begin{align}
\label{opt_lambda}
    \lambda^* \in \argmax_{\lambda \geq 0} \Pi_{i=1}^N p(l_i | \hat{\bm{\alpha}}_{\textsf{post}}(x_i;\lambda)) q(\lambda),
\end{align}
where $p(l |\hat{\bm{\alpha}}_{\textsf{post}}(x;\lambda)) = \int \pi_l(x) Dir(\bm{\pi}(x)|\hat{\bm{\alpha}}_{\textsf{post}}(x; \lambda)) \mathrm{d} \bm{\pi}(x)$ for any $l$ and $l_i= l$ is defined such that $\rho(y_i) \in L_l$. Intuitively, $\lambda^*$ adjusts the goodness-of-fit to the given data and the a priori requirement about conservativeness based on the amount of data.

\section{Theoretical Analysis for Estimation of Pseudo-Count Function}
\label{Analysis}
We will show that, under some conditions, the mean and the covariance matrix of the Dirichlet posterior with the parameter function in (\ref{alpha_hat_post}) converges to the true probability vector and zero-matrix at any $x \in \mathcal{X}$, respectively. 
This indicates that our solution to Problem 1 converges to a true solution asymptotically.

For a covariance kernel $g(\cdot, \cdot|\bm{\theta})$ of a GP,
we define the function space $\mathcal{A}$ induced by $g(\cdot, \cdot|\bm{\theta})$ as
\begin{align}
    \mathcal{A} = \{ \sum_{i=1}^k a_i g(x_i, \cdot|\bm{\theta}) \;|\; \bm{\theta} \in \mathrm{S}(H), k \geq 1, a_i \in \mathbb{R}, x_i \in \mathcal{X} \},
\end{align}
where $\mathrm{S}(H)$ is the support of the prior $H$ for $\bm{\theta}$. We denote $\bar{\mathcal{A}}$ as the supremum norm closure of $\mathcal{A}$. That is,
\begin{align}
    \bar{\mathcal{A}} = \{ f \;|\; N_\varepsilon(f) \cap \mathcal{A} \neq \emptyset, \forall \varepsilon>0 \},
\end{align}
where $N_\varepsilon(f) = \{ f': \mathcal{X} \to \mathbb{R} \;|\; \sup_{x \in \mathcal{X}}|f'(x) - f(x)| < \varepsilon \}$.
Denoted by $\mathcal{C}(\mathcal{X})$ is the set of continuous functions on $\mathcal{X}$.
\begin{theorem}
\label{thm_process_convergence}
    Consider $m$ LGP priors with the covariance kernels $\{ g_l \}_{l=1}^m$, input-trajectory data set $X=\{x_i\}_{i=1}^N$ and $Y=\{ y_i \}_{i=1}^N$, a robustness degree $\rho$, and disjoint intervals $L_1, \ldots, L_m$, where each $x \in X$ is i.i.d. from a density $p(x)$ and $\bm{\pi}(x)$ denotes the probability distribution of $\rho(y)$ given $x$ defined in (\ref{discretized_process}).
    Suppose that $p(x) > 0$ for any $x \in \mathcal{X}$ and all $g_l$ satisfies the following assumptions.
\begin{description}
    \item[(A1)] $\exists u, \exists v > 0$, $\forall x \in \mathbb{R}^d$, $v \leq g_l(x,x) \leq u$,
    \item[(A2)] $\exists C>0, \exists q > 0$, $[g_l(x,x) + g_l(x',x') - 2g_l(x,x')]^{1/2} \leq C||x-x'||^q$, $\forall x,x' \in \mathbb{R}^d$,
    \item[(A3)] $\forall N \geq 1, \forall x_1, \ldots, x_N \in \mathbb{R}^d$, $\mathrm{det} (G_l) \neq 0$, where $G_l =(g_l(x_i, x_j))_{i,j}$,
    \item[(A4)] each $\bar{\mathcal{A}}_l$ induced by $g_l$ is $\mathcal{C}(\mathcal{X})$.
\end{description} 
    Then, for any $x \in \mathcal{X}$, any $\bm{\alpha}_\textsf{prior} \in \mathbb{R}^m$, and any $\lambda \geq 0$, as $N \to \infty$, we have, w.p.1,
    \begin{align}
        \hat{\bm{\pi}}_E(x) &\to \bm{\pi}(x),\\
        \hat{\bm{\pi}}_V(x) &\to \bm{0},
    \end{align}
    where $\hat{\bm{\pi}}_E(x)$ and $\hat{\bm{\pi}}_V(x)$ are the mean and covariance matrix of $Dir(\cdot | \hat{\bm{\alpha}}_{\textsf{post}}(x; \lambda))$, respectively, where $\hat{\bm{\alpha}}_{\textsf{post}}(x; \lambda)$ is defined in (\ref{alpha_hat_post}).
    
\end{theorem}


To prove Theorem 1, we first introduce a weak neighborhood and the Kullback-Leibler (KL) support. Let $\mathcal{F}$ be the set of density functions over $\mathcal{X}$.
We define a weak neighborhood \cite{ramamoorthi2003bayesian} of a density function $p$ in $\mathcal{F}$ for any $n$ and any collection of $n$ continuous and bounded functions $\{ h_i : \mathcal{X} \to \mathbb{R} \}_{i=1}^n$ as
\begin{align}
    B(p, \varepsilon) = \{ q \in \mathcal{F} \;|\; \abs{ \int_\mathcal{X} h_i(x)q(x) \mathrm{d}x - \int_\mathcal{X} h_i(x)p(x) \mathrm{d}x } < \varepsilon, \nonumber \\ i = 1,\ldots, n \}.
\end{align}
Then, $p \in \mathcal{F}$ is said to be in the KL support \cite{tokdar2007posterior} of a prior $\Pi$ on $\mathcal{F}$ if, for any $\varepsilon > 0$, we have $\Pi(\{ q \in \mathcal{F} \;|\; D_{KL}(p||q) < \varepsilon \}) > 0$, where $D_{KL}(p||q) = \int p(x) \log{\frac{p(x)}{q(x)}} \mathrm{d}x$. We denote the KL support of $\Pi$ by $KL(\Pi) \subseteq \mathcal{F}$.
For any prior $\Pi$ on $\mathcal{F}$ and any $X$ consisting of i.i.d. samples from a density $p$, the posterior $\Pi(\cdot|X)$ is said to be weakly consistent at $p$ if, for any weak neighborhood $B$ of the probability density $p$, the posterior mass $\Pi(B | X)$ converges to $1$ almost surely as the number of samples $|X|$ goes to $\infty$.
Recall that $p_\mathrm{E}(x|l, X)$ and $p_\sigma(x|l, X)$ are the mean and the standard deviation of the LGP posterior conditioned on the input samples whose corresponding robustness values fall within $L_l$.

The following lemma indicates that $p_\mathrm{E}(x|l, X)$ and $p_\sigma(x|l, X)$ converge to the true density and $0$, respectively, if the covariance kernel $g_l$ of the LGP satisfies the regular conditions (A1)-(A4) and $\bar{\mathcal{A}}$ is $\mathcal{C}(\mathcal{X})$.

\begin{lemma}
\label{lemma_LGP_convergence}
Consider an LGP prior with the covariance kernel $g$ and a set of input samples $X \subset \mathcal{X}$, where each $x \in X$ is i.i.d. from a density $p(x)$. Suppose that $g$ satisfies the assumptions (A1)-(A4)
and $\bar{\mathcal{A}}$ induced by $g$ is $\mathcal{C}(\mathcal{X})$. Then, for any $x$ in $\mathcal{X}$, the mean $p_\mathrm{E}(x|X)$ and the standard deviation $p_\sigma(x|X)$ of the LGP posterior conditioned on $X$ converge to $p(x)$ and $0$ as $N \to \infty$, respectively.
\end{lemma}

\begin{proof}
    By Theorem 4.6 in \cite{tokdar2007posterior}, we have $p \in KL(\Pi)$, where $\Pi$ is the LGP prior. Thus, by Theorem 3.1 in \cite{tokdar2007posterior}, the posterior $\Pi(\cdot|X)$ is weakly consistent at $p$. Therefore, by Theorem 4.2.1 in \cite{ramamoorthi2003bayesian}, we conclude that $p_\mathrm{E}(x|X) \to p(x)$ and $p_\sigma(x|X) \to 0$ as $|X| \to \infty$.
\end{proof}

\begin{proof}[Proof of Theorem 1]
    For any $x$ in $\mathcal{X}$ and each $l$, $\hat{\alpha}_l(x ; \lambda) \geq 0$ for a large $N$ by Lemma \ref{lemma_LGP_convergence}. So, $\hat{\pi}_{\mathrm{E},l}(x)$ is written as
    \begin{align}
        \hat{\pi}_{\mathrm{E},l}(x) &= \frac{ \frac{N_l}{N} (p_{\mathrm{E}}(x|l, X) - \lambda p_{\sigma}(x|l, X)) + \frac{\alpha_{\textsf{prior},l}}{N} }{ \sum_i \left( \frac{N_i}{N} (p_{\mathrm{E}}(x|i, X) - \lambda p_{\sigma}(x|i, X)) + \frac{\alpha_{\textsf{prior},i}}{N} \right) }, \nonumber \\
        \intertext{where $N_l = |X_l|$, by the law of large numbers \cite{borovkov1999probability} and Lemma \ref{lemma_LGP_convergence}, we have, w.p.1, }
        \hat{\pi}_{\mathrm{E},l}(x) &\to \frac{p(l)p(x|l)}{p(x)} = \pi_l(x) \mbox{ as } N \to \infty.
    \end{align}
    For visibility, we denote $\hat{\alpha}_{\textsf{post},l}(x)$, $\sum_{i \neq l} \hat{\alpha}_{\textsf{post},i}(x)$, and $\sum_i \hat{\alpha}_{\textsf{post},i}(x)$ by $\alpha_{l,x}$, $\alpha_{-l,x}$, and $\alpha_{0,x}$, respectively.
    Then, with respect to $(i,j)$-th element $\hat{\pi}_{\mathrm{V}, ij}(x)$ of $\hat{\bm{\pi}}_\mathrm{V}(x)$, by the law of large numbers and Lemma \ref{lemma_LGP_convergence}, we have that, w.p.1,
    \begin{align}
        \hat{\pi}_{\mathrm{V}, ii}(x) &= \frac{ \alpha_{i,x}(\alpha_{0,x} - \alpha_{i,x}) }{ \alpha_{0,x}^2(\alpha_{0,x}+1) } \to 0 \mbox{ as } N \to \infty, \\
        \hat{\pi}_{\mathrm{V}, ij}(x) &= \frac{ -\alpha_{i,x} \alpha_{j,x} }{ \alpha_{0,x}^2(\alpha_{0,x}+1) } \to 0 \mbox{ as } N \to \infty.
    \end{align}
    
\end{proof}
Intuitively, by Theorem \ref{thm_process_convergence}, the estimator of the discrete random process and the posterior covariance matrix converge to the true one and the zero-matrix, respectively, when the total number of samples is large enough. This validates our proposed DLGP model.


\section{Example}
\label{Example}
\subsection{Setup}
\begin{figure}[thbp]
  	\centering
  	\subfigure[]{
  		\includegraphics[width=0.475\linewidth]{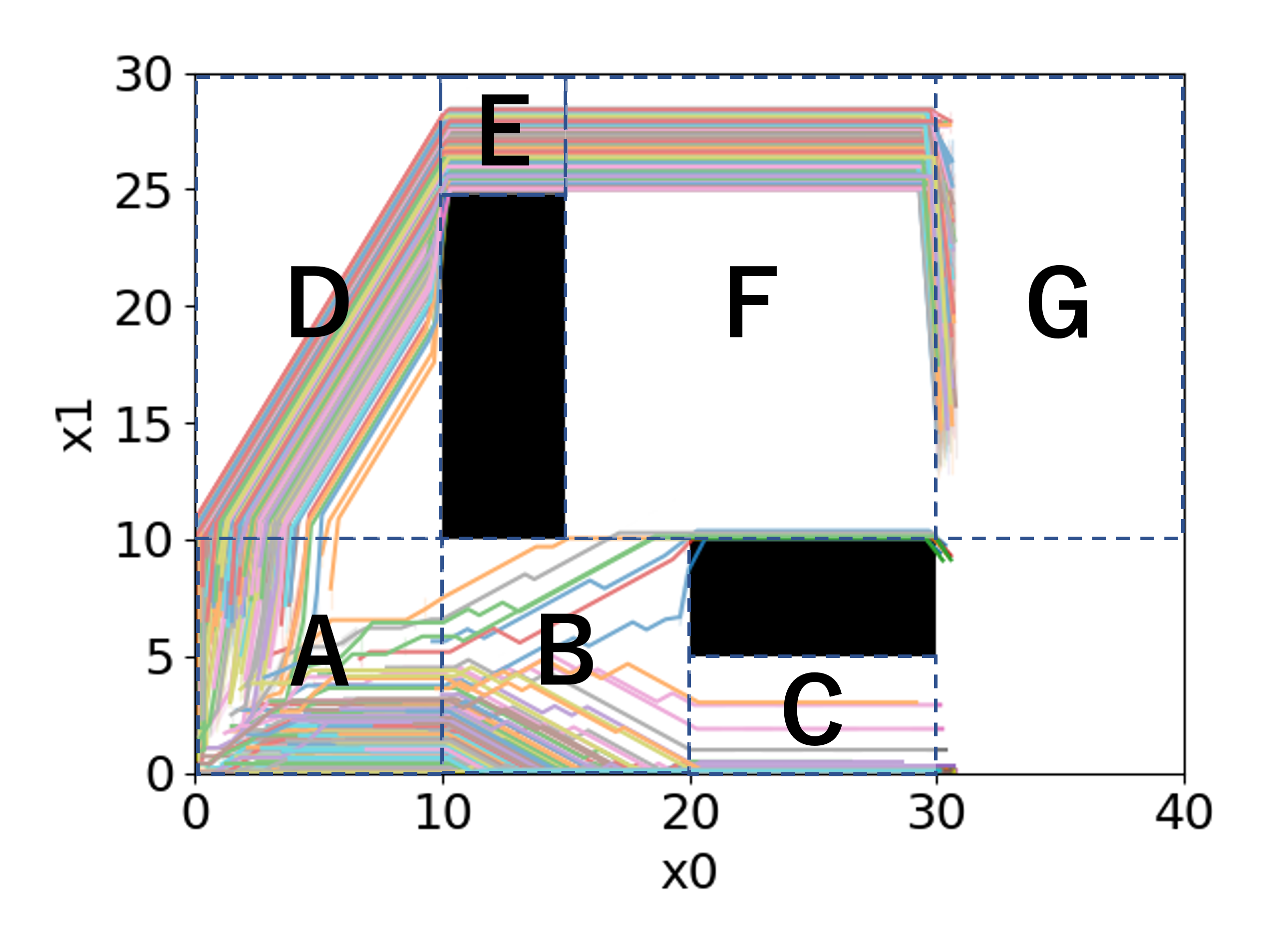}
            \label{ex:env_traj}
  	}
  	\subfigure[]{
  		\includegraphics[width=0.475\linewidth]{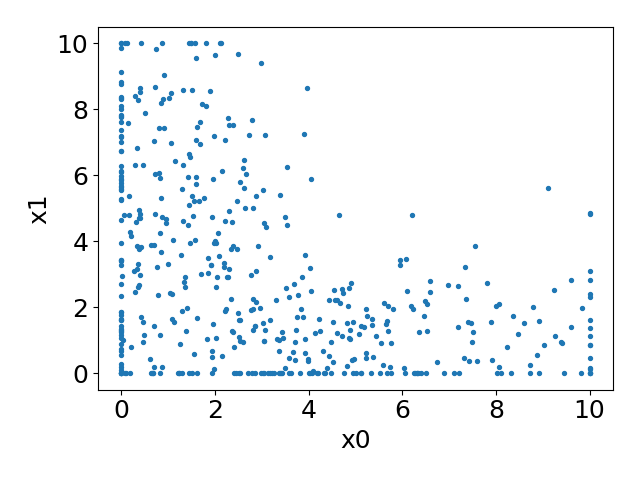}
  		\label{ex:input}
  	}
  	\caption{(a) An illustrative example of an environment with two obstacles (black rectangles) and yielded $500$ trajectories. The input region $\mathcal{X} = [0,10]^2$ is given as the set of initial starting locations of the robot. The goal location of the moving robot is given by $(35,5)$. (b) The $500$ inputs sampled in $\mathcal{X}$ through (\ref{sample_dist}).}
   \vspace{5mm}
\end{figure}

We applied the proposed method to a two-dimensional path planning problem of a robot in an environment with two obstacles depicted in Fig.\ref{ex:env_traj}.  The specification was ``the robot reaches the point $(35,5)$ within an error of 5 on each axis in $10$ seconds". So, by describing the specification using an STL formula \cite{fainekos2009robustness, donze2010robust}, the corresponding robustness degree was given by $\rho(y) = \max_{t \in [0,10]} (5 - \max{ \{|35 - y_0(t)|, |5 - y_1(t)|\} })$, where $y(t)=(y_0(t),y_1(t))$ in $\mathbb{R}^2$ is the location of the robot at the time $t$. We considered the set of initial locations as $\mathcal{X} = [0, 10]^2$ and denoted the location as $x = (x_0, x_1)$.
The robot transited per $0.1$ second. The transition distance $v$ was determined as $0.3 + \varepsilon$ regardless of the location, where $\varepsilon$ followed the uniform distribution over $[0, 0.5]$. When the robot was about to hit the wall and the obstacles under the decided action, it avoided colliding with the wall and reselected the nearest moving direction.
We summarize the detailed behavior in each area in Table \ref{table:robot_direction}, where $\sigma(x) = 1/(1+\exp(x))$. Each area is depicted in Fig. \ref{ex:env_traj}. The robot stayed at the current location with the indicated probabilities in Table \ref{table:robot_direction} at any time step when it was in $\mathrm{E} = [10, 15] \times [25, 30]$ and $\mathrm{C}= [20, 30] \times [0, 5]$.

\begin{table}[thbp]
  \caption{The directions the robot moves to and the corresponding probabilities. The indices denote the areas in the environment.}
  \label{table:robot_direction}
  \centering
    \begin{tabular}{ c||c c c c} \hline
       & direction & probability \\ \hline \hline
      A & upward / right & $\sigma(x_0 - x_1)$ / $1 - \sigma(x_0 - x_1)$ \\ 
      B & \begin{tabular}{c}upper / lower right\\ tilted by 60 degrees\end{tabular} & $\sigma(5-x_1)$ / $1 - \sigma(5-x_1)$ \\ 
      C & right / stay & $0.3$ / $0.7$  \\ 
      D & upper right tilted by $30$ degrees & $1$  \\ 
      E & right / stay & $0.9$ / $0.1$  \\ 
      F & right & $1$  \\
      G & downward & $1$ \\ \hline
    \end{tabular}
  \end{table}
  

Shown in Fig.\ref{ex:input}, we generated $N=500$ inputs $X$ from $\mathcal{X}$. Then, trajectories $Y$ were obtained by executing 10 seconds per the input with the above dynamics. The $N$ initial locations of the robot was randomly determined as follows:
\begin{align}
\label{sample_dist}
    & x = ( \min \{ \max\{0, x'_0\}, 10 \}, \min \{ \max\{0, x'_1\}, 10 \}), \nonumber \\
    & x^\prime \sim 0.5 \mathcal{N}((1,5)^\top, \mathrm{diag}(2,10)) + 0.5 \mathcal{N}((5,1)^\top, \mathrm{diag}(10,2)).
\end{align}

We set the robustness levels as $L_1 = (-\infty, -10]$, $L_2=(-10, 0]$, and $L_3 = (0, \infty)$ with $m=3$. We adopted Gaussian kernels for $m$ LGPs as $g_l(x,x') = \theta_{l,1} \exp{(\theta_{l,2}||x-x'||^2)}$. Furthermore, we set the hyper-priors for $\theta_{l,1}$ and $\theta_{l,2}$ as the half-normal distributions with the standard deviation $1$, respectively. We computed each posterior through (\ref{LGP_posterior}) and (\ref{hyper_param_posterior}) using the no-U-turn sampler \cite{hoffman2014no} by dividing the set of input locations into grids with the grid width $0.5$.
We placed a Gamma prior whose mode is $2$ and variance is $3$ for $\lambda$.
We chose a non-informative prior parameter vector of the Dirichlet distribution as $\bm{\alpha}_{\textsf{prior}} = (1/3, 1/3, 1/3)$. 
To confirm the effectiveness of the LGP confidence, we considered the three versions of the estimators with $\lambda$ in $\{0, \lambda^*, 1\}$, where $\lambda^*$ is obtained from (\ref{opt_lambda}).

We compared our proposed method with the following two estimation methods. Both methods estimated the discrete probability in (\ref{discretized_process}) and quantify the uncertainty of the estimation using the input samples labeled as $\{ (x_i,l) \;|\; x_i \in X, y_i \in Y, \rho(y_i) \in L_l \}$. The first one is the version of \cite{charpentier2020posterior} using kernel density estimators (KDEs) \cite{scott2015multivariate} instead of normalizing flows for a fair comparison under small data. We call the method \textsf{DKDE} in Example. A KDE is a well-known density estimator in a frequentist manner. We used the Gaussian kernel for each KDE and tuned its hyper-parameters by Scott's rule \cite{scott2015multivariate}.
The second one called \textit{Gaussian Dirichlet process} (\textsf{GDP}) \cite{milios2018dirichlet} is a Gaussian process classification method. This utilizes a Dirichlet-categorical model and approximates the $m$-dimensional Dirichlet distribution at each $x$ with $m$ independent log-normal distributions. The mean parameter of $l$-th log-normal distribution was regressed by the GP with a Gaussian kernel.
We conducted each method $20$ times for the obtained input-trajectory data.

\subsection{Result}
\begin{figure}[thbp]
 \centering
  	\subfigure[]{
  		\includegraphics[width = 4.12cm]{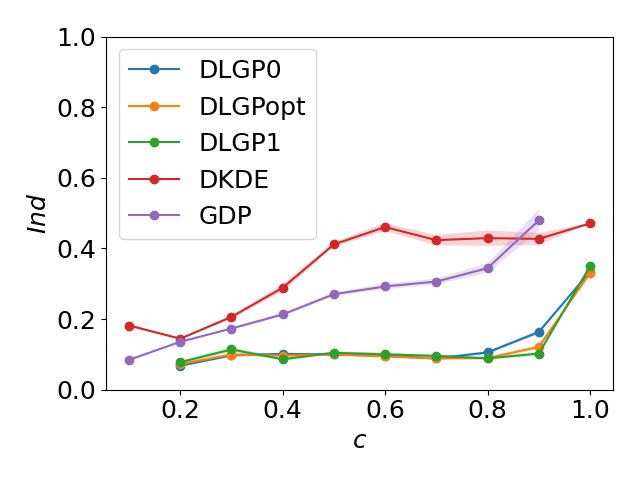}
            \label{ex:est_error}
  	}
  	\subfigure[]{
  		\includegraphics[width=4.12cm]{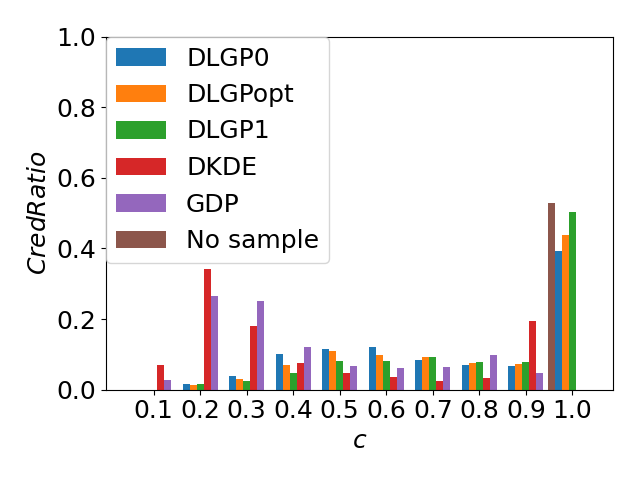}
  		\label{ex:cred_ratio}
  	}
 \caption{(a) Mean of the index $Ind(c)$ in (\ref{index}) for $20$ estimations. Each shaded region represents the standard deviation. We omit plots for which the corresponding $\mathcal{X}(c)$ are empty. (b) Mean of the index $CredRatio(c)$ in (\ref{cred_ratio}) for $20$ estimations. \textsf{No sample} at $c=1$ represents the value of $CredRatio(c)$ over the area where no sample is obtained.}
\vspace{5mm}
\end{figure}
We define the confidence band function $\hat{\pi}_C: \mathcal{X} \to [0,1]$ as $\hat{\pi}_C(x) = \sum_{l=1}^m (\hat{\pi}_{\mathrm{U}, l}(x) - \hat{\pi}_{\mathrm{L},l}(x))/m$, where $\hat{\pi}_{\mathrm{U},l}$ and $\hat{\pi}_{\mathrm{L},l}$ are the lower and upper confidence bounds defined such that
\begin{align}
\label{upper_confidence}
\Pr(\hat{\pi}_l(x) \geq \pi_{U,l}(x) \;|\; x) &= \frac{\beta}{2}, \\
\label{lower_confidence}
\Pr(\hat{\pi}_l(x) \leq \pi_{L,l}(x) \;|\; x) &= \frac{\beta}{2},
\end{align}
where $\beta=0.05$.
Intuitively, for each $l$ and $x$, $\hat{\pi}_{\mathrm{U},l}(x)$ and $\hat{\pi}_{\mathrm{L},l}(x)$, respectively, represent the maximum and minimum values that $l$-th component of $ \hat{\bm{\pi}}(x) \sim Dir(\cdot|\hat{\bm{\alpha}}_\textsf{post}(x))$ can take with probability (confidence) $1-\beta$. So, when $\hat{\pi}_C(x)$ is large (resp. small), the uncertainty of estimation is large (resp. small). In the following, we evaluate the confidence band value by dividing it into 10 equal parts on $[0, 1]$.

To show that the proposed method achieves a small error between the estimated discrete random process and the true one for robustness levels, we introduce the following index:
\begin{align}
\label{index}
    Ind(c) \! = \! \frac{\int_{x \in \mathcal{X}(c)} \sum_{l=1}^m (\hat{\pi}_{\mathrm{E},l}(x) - \pi_{l}(x))^2 \mathrm{d}x} {\int_{x \in \mathcal{X}(c)} \mathrm{d}x}, 
\end{align}
where $\mathcal{X}(c) = \{ x \in \mathcal{X} \;|\; \hat{\pi}_C(x) \in ( c - 0.1, c]$, recall that $\hat{\bm{\pi}}_{E}(x) = \mathbb{E}[\hat{\bm{\pi}}(x) \;|\; x ]$, and $\bm{\pi}$ is obtained as the proxy of the true one by the following procedure. First, we generate $10^5$ initial locations uniformly and at random and obtain $10^5$ trajectories. Then, we calculate the ratio of the pseudo-counts defined in (\ref{ideal_pseudo_count}) through $3$ KDEs with the bandwidth $0.01$.
Intuitively, the index $Ind(c)$ represents how close the estimated probability vector is to the true one on the region of $\mathcal{X}$ where the confidence band is in $(c-0.1, c]$. 

Moreover, to show the effectiveness of the conservative evaluation of the pseudo-count for the estimation confidence on the region where no sample is obtained, we define the following index:
\begin{align}
\label{cred_ratio}
    CredRatio(c) = \frac{ \int_{x \in \mathcal{X}(c)} \mathrm{d}x }{ \int_{x \in \mathcal{X}} \mathrm{d}x }.
\end{align}
Intuitively, $CredRatio(c)$ denotes the ratio of the region whose confidence band is in $(c -0.1, c]$.

 Fig. \ref{ex:est_error} denotes the means and standard deviations of the estimation error $Ind$ for each method. We denote the results obtained from our proposed method with $\lambda=0, \lambda^*, 1$ as \textsf{DLGP0}, \textsf{DLGPopt}, and \textsf{DLGP1}, respectively. We observe that our proposed method \textsf{DLGP} achieves the lowest estimation error for all confidence bands and the estimation errors when $\lambda = 0$ and $\lambda^*$ are almost increasing for the confidence band. This implies that \textsf{DLGP} most accurately captures the discrete process for the robustness levels. On the other hand, \textsf{DKDE} has a high estimation error for each confidence band. This seems to be due to the difficulty of determining the appropriate bandwidth of the kernel to obtain plausible pseudo-counts. Shown in Fig. \ref{ex:cred_ratio} are the values of $CredRatio$ for each method and the ratio of the area without samples to $\mathcal{X}$ when dividing $\mathcal{X}$ into grids with the grid width $0.5$. Note that the values of $\hat{\pi}_C$ in the no-sample areas should be close to $1$ since the confidence of the estimation should be small in those areas. Fig. \ref{ex:cred_ratio} indicates that $\lambda = \lambda^*$ and $1$ more accurately capture the no-sample area than other methods through the high values of confidence bands. This seems to be attributed to their conservative pseudo-count evaluation. We observed that \textsf{GDP} is over-confident compared with \textsf{DLGP}. This seems to be because the uncertainty of \textsf{GDP} is obtained only from the GP posteriors instead of the Dirichlet distributional uncertainty based on observation counts.

\section{Conclusion}
In this paper, for a black-box stochastic system that generates a trajectory under an input equipped with a robustness degree (real-valued performance measure for a trajectory) that describes a requirement, we considered the problem of estimating the probability distribution of the robustness value for any input on a compact region. We relaxed the problem into estimating the discrete random process that represents the probability of the robustness value classified into each disjoint real number interval. Then, based on a Dirichlet random field over the input region and multiple LGPs, we proposed a semiparametric Bayesian method to estimate the discrete process with conservative confidence bounds.
We showed that the proposed Bayesian model converges to the true discrete random process under some mild conditions.  
Future works are twofold: 1) Combining the proposed method with active testing. 2) Dealing with the high dimensional input space.


\bibliography{reference}

\end{document}